\documentclass{sig-alternate}

\usepackage[utf8x]{inputenc}

\usepackage{url}
\usepackage{algorithm}
\usepackage{algorithmic}

\newcommand{\hte}[1]{\mbox{HybridTE$_{#1}$}}

\begin{document}

\title{HybridTE: Traffic Engineering for Very Low-Cost Software-Defined Data-Center Networks}

\numberofauthors{2} 
%
\author{
%
%
\alignauthor
Philip Wette\\
       \affaddr{University of Paderborn}\\
       \affaddr{Warburger Stra{\ss}e 100}\\
       \affaddr{33098 Paderborn, Germany}\\
       \email{philip.wette@uni-paderborn.de}
\alignauthor
Holger Karl\\
       \affaddr{University of Paderborn}\\
       \affaddr{Warburger Stra{\ss}e 100}\\
       \affaddr{33098 Paderborn, Germany}\\
       \email{holger.karl@uni-paderborn.de}
}
\date{26 March 2013}

\maketitle
\begin{abstract}
	The size of modern data centers is constantly increasing. 
	As it is not economic to interconnect all machines in the data center using a full-bisection-bandwidth
	network, techniques have to be developed to increase the efficiency of data-center networks. 
	The Software-Defined Network paradigm opened the door for centralized traffic engineering (TE) in such environments.
	Up to now, there were already a number of TE proposals for SDN-controlled data centers that all work very well.	
	However, these techniques either 
	use a high amount of flow table entries or a high flow installation rate that overwhelms available switching hardware, or 
	they require custom or very expensive end-of-line equipment to be usable in practice.
	
	We present HybridTE, a TE technique that uses (uncertain) information about large flows. Using this extra information, our technique 
	has very low hardware requirements while maintaining better performance than existing TE techniques.
	This enables us to build very low-cost, high performance data-center networks.
	
\end{abstract}

\category{C.2.1}{Network Architecture and Design}[Network communications, Packet-switching networks]
\category{\\C.2.4}{Distributed Systems}[Network operating systems]

\terms{Software-Defined Networks, packet sampling, traffic engineering}


\section{Introduction}
	To increase the throughput of data-center networks a lot of traffic engineering (TE) techniques have been proposed.
	Some of them are distributed in nature, such as Equal-Cost Multi-Path (ECMP), and some of them are using the novel 
	Software-Defined Network (SDN) paradigm, 
	to make fine grained routing decisions on a flow-based level by using global information.

	Recent studies of data-center traffic \cite{datacentersInTheWild, MSR-datacenters} revealed that the traffic in data centers is very diverse. 
	80\,\% of all flows are shorter than 10\,KB, whereas more than half of
	all transported bytes reside in only a few flows. These flows are called \emph{elephants} (as opposed to \emph{mice}).

	In ECMP, each flow (both elephants and mice) is randomly assigned to one of multiple shortest paths between the source and destination;
	ECMP is the de facto standard in traditional data centers and aims at evenly distributing load to all links. If, however, multiple elephants are 
	assigned to a shared link, these elephants are competing for
	data rate although there might be paths with free data rate left.
	This is where most novel TE techniques \cite{microte, hedera, devoflow} come into play. They detect colliding elephants and reroute them  
	based on global information about the
	network.
	
	According to \cite{datacentersInTheWild, MSR-datacenters}, the average flow size in contemporary data centers is approx. 146\,KB. 
	Thus, to saturate a full duplex 10\,Gbps link, it requires approx.\,20,000 new flows every second.
	All available SDN TE techniques either require one flow entry per flow (i.e., they do not use wildcard flows) or
	switches with custom logic which can not be bought off-the-shelf.
	Without using custom logic, a 48-port switch requires 480K flow table entries on average when these TE techniques are used.
	Although there exist SDN-capable switches with such large flow tables, these switches \emph{cannot} keep up with the high flow arrival rate.
	NoviFlow's NoviSwitch, for example, supports up to 1\,million flow entries but it is only able to process 2500 flow installations per second,
	which is two orders of magnitude less than what is required to saturate all 48 ports.
	
	Our contributions are as follows:
	We propose HybridTE, a novel TE technique for SDN-controlled data-center networks that uses information about elephant flows. 
	In turn it only requires 
	very few flow entries in the switches and has a very low flow installation rate.
	HybridTE allows us to build high-performance data-center networks using low-cost off-the-shelf SDN switches that need only limited space and 
	processing capabilities without loosing network performance.
	We show the performance of HybridTE using a MaxiNet-based network emulation with realistic data-center traffic and compare our results to
	traditional ECMP and to Hedera \cite{hedera}. 
	The key idea of HybridTE is to handle mice differently than elephants.
	To this end, we require explicit knowledge about elephants which can be created, for example, by packet-sampling techniques.
	Different elephant-detection techniques create different qualities of information. 
	In a hypothetical ideal case, elephant detection
	reports all existing elephants without any delays.
	In reality, however, elephants will be detected some time after they start and with some amount of false positives and false negatives.
	
	We show how the different levels of false positives, false negatives and delay impact the quality of our traffic engineering scheme.	
	This makes this paper the first to show the relation between the amount of explicit information and the performance gained by this information
	in a realistic data-center environment.
	Data shows that even with 50\,\% false negatives and a reporting delay of 1\,s, the performance of HybridTE is comparable with Hedera and ECMP.
	Also, the amount of false positives does not have any significant impact on our proposed scheme.

	The rest of this paper is structured as follows: Section~\ref{sec:relatedWork} gives an overview of related work.
	Section~\ref{sec:hybridte} motivates the problem addressed by this paper and introduces its solution: HybridTE, a novel routing algorithm for software-defined
	data-center networks. This algorithm is evaluated using the network emulator MaxiNet \cite{wette14b} under realistic traffic in Section~\ref{sec:evaluation}. Section~\ref{sec:conclusion} concludes
	this paper.

\section{Related Work}
\label{sec:relatedWork}
To speed up applications running in the data center, various techniques have been proposed recently. 
The research area most related to HybridTE
is \emph{traffic engineering} (TE).
TE focuses on the routing of individual flows.
In the traditional sense, it works on time scales of hours to days. Using SDN, however, traffic engineering can be used
on a second or even sub-second scale.

Hedera \cite{hedera} was one of the first TE schemes proposed for the SDN protocol OpenFlow.
Hedera uses a centralized scheduler to dynamically compute new routes for elephant flows. To identify elephant flows, 
the scheduler periodically polls data about all installed flows from all switches. 
Unfortunately, the centralized control of Hedera does not keep up with the high flow arrival rate of real data centers.

Just like Hedera, DevoFlow \cite{devoflow} uses a centralized scheduler to do TE. To remove load from this controller, DevoFlow
uses custom switching hardware and a modified version of the OpenFlow protocol. The evaluation with realistic data-center traffic, however, 
could not show any significant performance gain over ECMP.

MicroTE \cite{microte} uses short-term predictions of the traffic matrix to perform TE in data centers. Evaluation results show
that MicroTE performs well under realistic traffic assumptions. To predict traffic, MicroTE relies on modified end hosts to collect
server-to-server traffic matrices for sub-second time scales. As this does not scale with increasing number of servers, MicroTE is able to fall back to 
rack-to-rack traffic matrices. However, evaluation shows that in rack-to-rack mode MictoTE does not outperform ECMP.

CONGA \cite{conga} is a fully distributed congestion-aware load balancer for data centers which requires hardware support.
It has been implemented in custom ASICs and will be available soon in Cicso's top-of-the-line data-center switches.

As already stated in the introduction, we want to leverage low-cost off-the-shelf SDN switches for building data centers. Therefor, 
HybridTE is designed with the focus on low resource usage and compatibility with the OpenFlow standard.
To the best of our knowledge, none of the existing techniques complies with these requirements.

\section{HybridTE}
\label{sec:hybridte}
\subsection{Problem}

	Efficiently handling data-center traffic using SDN is complicated due to the limited scalability of the centralized SDN approach.
	The traffic patterns in data centers are changing very frequently \cite{microte} and traffic mostly consists of very small flows
	(90\,\% < 1\,MByte), whereas most bytes are transported by a minority of elephant flows.
	The aggregated flow arrival rate is too high to be handled in a centralized manner and most of the decisions (i.e., for mice) 
	have only very little influence at all on the overall	load situation.
	
	Thus, when using SDN in the data center, it does not pay off to make individual routing decisions for each single flow.
	Even if we did, it would lead to performance penalties. 
	10\,KB sent over a 10\,Gpbs line takes 1\,µs, which means that even if decision making only takes an additional 1\,µs, 
	the flow finishing time is doubled. 
	
	For a TE scheme to be usable in practice, the requirements on the switching hardware should be as low as possible which is why 
	HybridTE uses only few wildcarded and exact-match flow-table entries while having a very low flow installation rate.
	
\subsection{Idea}
	To a) remove load from the central controller and b) reduce delay in handling mice flows, HybridTE handles mice flows locally at 
	the switches by providing proactively installed static routes.
	When handling elephant flows using local decisions as well,
	this leads to avoidable congestion whenever two or more elephants are assigned to the same link while alternative routes
	with free capacity exist that could have been chosen using global knowledge.
	To this end, HybridTE requires knowledge about elephant flows. Once HybridTE is informed about an arriving elephant, this flow is routed individually
	using global knowledge.

	In this work, we study this question: Assuming we know which flows are becoming elephants during their lifetime, is it sufficient to have a very simple
	proactive routing scheme when performing traffic engineering on the elephants only?
	As in reality there is no explicit knowledge about \emph{all} elephants available, we determine 
	what number of elephants leads to which performance level, and how many elephants we have to be aware of to be at least as good as existing TE techniques.

\subsection{Static Routing}
	
\begin{figure}
		\begin{center}
			\includegraphics[scale=0.7]{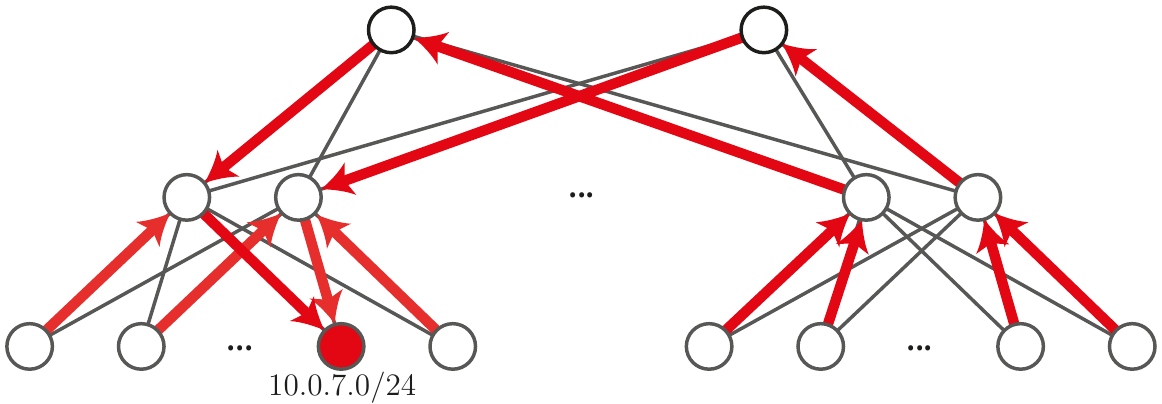}
		\end{center}
		\caption{Clos-like data-center topology with an exemplary forwarding tree for subnet 10.0.7.0/24.}
		\label{fig:trails}
	\end{figure}

	The static routing component in HybridTE provides basic connectivity and aims to spread traffic evenly over all links in the network and flow table entries
	evenly over all switches.
	To provide low latency while using as few resources as possible, the scheme forwards traffic along shortest paths only.
	Therefore, the scheme constructs one forwarding tree per destination rack.
	Figure~\ref{fig:trails} shows a typical data-center architecture consisting of a core layer (top), a pod layer (middle) and
	a layer of top-of-rack (ToR) switches. Each ToR switch connects one rack of servers (typically 20 to 40 servers per rack).
	ToRs are grouped within pods. Each pod has two pod switches which are interconnected through the core switches.
	We assume that within each rack, servers use a unique IP subnet. 
	This assumption allows us to implement one forwarding tree using one OpenFlow wildcard flow-table entry per switch only.
	In Section~\ref{sec:labelrouting} we show that when 
	using label routing techniques this assumption need not hold for HybridTE to work.
	
	HybridTE uses Algorithm~\ref{algo:staticroutes} to install static routes.
	In the algorithm, $R$ is the set of all racks and $\mathcal{S}_{i,j}$, $i,j \in R$, is the set of all shortest paths between racks $i$ and $j$.
	To create a forwarding-tree for rack $i \in R$, for each rack $j \in R$, $i \neq j$, one path $p \in \mathcal{S}_{i,j}$ is chosen uniformly at random. 
	Then, on each switch $s$ on the path $p$ that has not yet a rule to handle traffic destined to $i$,
	one wildcard rule is installed to forward all traffic destined to $i$ to the next switch on the path $p$.
	Note that this algorithm creates one forwarding \emph{tree} per destination.
	Figure~\ref{fig:trails} gives an example for such a tree.
	
\begin{algorithm}
\caption{Install Static Routes}
\label{alg1}
\begin{algorithmic}
\FOR{$(i,j) \in R \times R$}
		\STATE choose Path $p \in \mathcal{S}_{i,j}$ uniformly at random
		\FOR{each Switch $s $ on $p$}
			\IF{$s$ has no entry matching destination $i$}
				\STATE on $s$, install a wildcard entry matching all packets destined to $i$ with the action to forward the packet to the next switch on $p$
			\ENDIF
		\ENDFOR

\ENDFOR
\end{algorithmic}
\label{algo:staticroutes}
\end{algorithm}
	
	With these default routes in place, HybridTE creates basic connectivity and already spreads traffic over the whole topology.
	When using the OpenFlow protocol, this only requires
	$|R| + M$ wildcard flow entries at each ToR switch and $|R|$ entries at each core and pod switch where $M$ is the number of machines per rack.

\subsection{Support for Unstructured IP Addresses}
\label{sec:labelrouting}
	Using novel label switching techniques \cite{schwabe14, shadowMac}, the assumption about structured IP addresses made in the previous section can be dropped
	while keeping the same number of OpenFlow flow table entries at each switch. 
	This allows for endhost mobility without having to change IP addresses after moving a machine from one rack to another,
	which is a very important feature for upcoming cloud data centers.
	
	The core idea in \cite{schwabe14} is using spoofed ARP replies to create structure in the MAC addresses. 
	Whenever a host sends an ARP request for a host residing in rack
	$i$, the SDN controller intercepts the request and answers with a faked MAC address which uses the first bytes of the MAC address to encode the rack ID $i$
	and the last bytes to encode the endhost.
	In addition, on the ToR of rack $i$, a flow entry is installed to rewrite the faked MAC address back to the host's original MAC before delivering any packet. 
	Whenever a host moves from one rack to another, gracious ARP is used to promote the new MAC address of the host. To deliver any packets from flows
	to a recently moved host (which are addressed to the now outdated MAC), we install a rule to rewrite the MAC accordingly at the ToR of rack $i$.
	
	As the rack ID is encoded in the first bytes of the MAC addresses, the static routing component of HybridTE can use these bytes for wildcard routing
	instead of the IP subnet as before.
	Note that this does not change the number of required OpenFlow flow table entries at the switches.
	
\subsection{Reactive Elephant Routing}
	Using the static routing scheme only, congestion is very likely to appear. Assume we have two elephant flows between the same
	pair of ToR switches but between distinct hosts. As we are using a forwarding tree to route traffic, both elephants are sharing the available data rate 
	on the path while
	there could be another path with free residual data rate.

	To shift traffic from heavily used links to unused links, we reroute elephants off the static routes onto individually computed routes.
	To this end, we need to know which flow becomes an elephant in the first place. 
	As HybridTE uses wildcard flow entries to route traffic, we do not know which flows are hiding behind these rules and thus depend on 
	external elephant detection techniques \cite{wette13c}.
	In the scope of this paper, we do not care where this information comes from; there are a lot of different possibilities to detect elephants.
	More details are given in Section~\ref{sec:isElephantDetectionRealistic}.
	
	Whenever a new elephant between ToRs $i$ and $j$ is reported to HybridTE, this elephant is routed individually through the network.
	The new route can be computed by different rerouting algorithms that can make use of all information present at the 
	SDN controller.
	For simplicity, HybridTE reroutes the elephant to the shortest path $p$ between $i$ and $j$ which is least loaded.
	Rerouting is done by installing exact matching flow entries on each switch on the new path $p$.
	
\subsection{Periodic Elephant Rerouting}
	
	As old flows complete and new flows arrive, the load situation of the network is constantly changing. To account for this, HybridTE
	periodically reroutes all known active elephants.
	Most traffic is transported using TCP. As rerouting a TCP flow introduces packet reordering and changes the round trip time perceived by the flow, 
	rerouting has a negative influence on the flow's data rate, which is why flows should not be rerouted too often.
	We decided to reroute every 5\,s which is  in line with what is discussed in \cite{hedera}.

	To receive a list of all reported but still active elephant flows, all core and pod switches are queried for their installed
	exact matching flow table entries.
	Using this information, we can compute the average data rate of each elephant and the path over which each elephant is 
	routed.\footnote{There are different ways to get the routes of all reported still active elephant flows. We decided for this soft-state solution
	because it is simpler to implement and allows for very simple and fast fail-overs in case the HybridTE controller fails. }
	As we are only installing elephant flows as exact matching flows and only a few elephants exist concurrently, this will not become a performance bottleneck.
	Next, all switches are queried for their number of bytes transferred over each physical switch port. By record keeping, 
	HybridTE computes the average data rates of all links over the last 5 seconds.
	By subtracting the average data rates of all elephants from the link data rates we can express the
	data rate on the links consumed by mice only; this can be seen as the background traffic for rerouting elephants.
	
	Finding an optimal routing for the active elephants is an $\mathcal{NP}$-hard problem.
	Just like Hedera, we use the \emph{Global First Fit} \cite{hedera} heuristic to solve it.	
	Global First Fit starts by computing the \emph{natural demand} of all elephant flows. The natural demand is the data rate the flow would achieve when
	the only bottleneck in the network was the host's network interfaces and all the rest of the network is non-blocking.
	Based on the link data rates we computed before, Global First Fit reroutes elephants by subsequently finding non-blocking paths
	for each elephant. For details on Global First Fit, see \cite{hedera}.

\subsection{Is Elephant Detection Realistic?}
\label{sec:isElephantDetectionRealistic}
	HybridTE leverages information about elephant flows in the network. Our evaluation shows (unsurprisingly) that with more accurate information, the 
	performance of HybridTE increases (Section~\ref{sec:results}). 
	But to which extent is information about elephant flows available in typical data centers?
	
	There are two different types of elephant detectors: the invasive type and the noninvasive type.
	The invasive type requires changes to the software running in the data center. 
	With the application information it can be told which
	data transfer is an actual elephant. Imagine a file download from an HTTP server.
	As the server knows the file size in advance, this information can directly be passed to HybridTE.
	
	The noninvasive type of elephant detectors does not require any changes to software and is much easier to deploy.
	For example, HadoopWatch \cite{HadoopWatch} monitors the log files of Hadoop for upcoming file transfers. 
	This way, the traffic demand of the Hadoop nodes
	can be forecast with nearly 100\% accuracy and ahead of time.
	Packet sampling is another noninvasive technique to identify elephants that is independent of the software running in the data center. 
	Using middleboxes or built-in features of switches, packet samples can periodically be taken and analyzed.
	Choi~et~al.\,\cite{Adaptivepacketsampling} showed that using a reasonable number of samples, elephants can be identified quickly with very low error.
	
	Different elephant detection techniques have different characteristics. HadoopWatch, for example, is able to predict elephants ahead of time, while 
	packet sampling usually takes more time to identify a flow as an elephant. In addition, different techniques yield different numbers
	of false positives, i.e., mice that are labeled as elephants by mistake. On the other hand, some elephants will not be detected at all.
	We give more insight into the effects caused by various values of false positives, false negatives and delays in Section~\ref{sec:experiments}.

\section{Evaluation}
\label{sec:evaluation}

\subsection{Emulation Environment}

In contrast to most other work in the context of traffic engineering in data-center networks, we are using highly realistic
traffic patterns for evaluation. Traffic is generated by DCT$^2$Gen \cite{wette14c}, a novel generator that uses
the findings of two recent studies \cite{datacentersInTheWild, MSR-datacenters} about traffic patterns in data centers.

We use MaxiNet \cite{wette14b} to emulate a mid-sized data center. MaxiNet is a distributed Mininet version that uses multiple
physical machines to emulate a network.
To support emulation of a 10\,Gpbs data-center network we rely on the concept of time dilation \cite{gupta2005infinity}, which basically means
slowing down the time by a certain factor (called time dilation factor). For our experiments, we used a time dilation factor of 150, meaning
each 10\,Gbps link was emulated by a 66.6\,Mbps link; in turn, the overall running time of each experiment increased by a factor of 150.

For evaluation, we implemented HybridTE, ECMP and Hedera on top of the OpenFlow controller Beacon \cite{beacon13} using OpenFlow\,1.0.
Both the connection between the switches and the controller and the controller itself were 
\emph{not} subject to time dilation, leading to nearly instant decision making.
This removes any effects possibly resulting from a bad controller placement or hardware too slow to host the controller.

The physical machines used to compute the emulation are four Intel i7\,960 CPUs with 24\,GB of memory and
five \mbox{Intel} i5\,4690 CPUs with 16\,GB of memory. The OpenFlow controller is hosted at an \mbox{Intel} Core\,2\,Duo\,E8400 running at 3\,Ghz
with 8\,GB of memory. All physical machines are wired in a star topology with 1\,Gbps Ethernet.

\subsection{Experiments}
\label{sec:experiments}
To find out how HybridTE performs, we emulate 60 seconds of data-center traffic on a Clos-like, full duplex 10\,Gbps network consisting
of 1440 hosts organized in 72 racks. During that time, approximately six million Layer~2 flows are emulated.
Each pod consists of eight racks and each pod has two pod switches. The core of the network consists
of two switches.
Due to the used time dilation factor of 150, one experiment takes two and a half hours walltime plus time for setting up the emulation and
compressing and storing the results.

The metric used to compare the different routing algorithms is the \emph{average flow completion time}, i.e., the average time 
between the first packet of a flow is sent until the last packet of that flow is received.
As argued in \cite{varys,dukkipati2006flow}, the flow completion time is one of the most important metrics for data centers running multi-staged applications where one
stage can only start when the prior stage is completed.

As already argued in Section~\ref{sec:isElephantDetectionRealistic}, different elephant detection techniques have different properties.
To account for this, we evaluate HybridTE under different numbers of false negatives, false positives and reporting delays.

\subsubsection{False Negatives: Actual elephants not reported}
\label{sec:results}

To see how HybridTE performs under different rates of false negatives, we evaluated it with information about
a) 100\,\%,
b) 75\,\%, and
c) 50\,\% of all elephants.
In the following, we reference these cases as \hte{100}, \hte{75}, and \hte{50}, respectively.
These correspond to 0\,\%, 25\,\%, and 50\,\% false negatives.
False negatives were drawn uniformly at random from all elephants.
We compare the results of HybridTE against ECMP and Hedera under four different load levels ($\times 1$, $\times 1.5$, $\times 1.75$, and $\times 2$).
A load level was created by scaling down the data rates of the emulated links while keeping the same emulated NIC speeds.
We used DCT$^2$Gen to generate ten independent \emph{flow traces} describing which host sends how much payload via TCP to which other host at which time.
For each load level, we repeated the experiment with each of the flow traces, which resulted in 200 experiments.
Including all overheads, the experiments took more than one month to complete.

As data-center traffic follows heavy-tailed distributions, 
the ten experiments we conducted for each configuration are not enough to tell the
statistically correct mean. This is why in the following we show the results for each run of the experiment individually.
We defined an elephant to be a flow transmitting more than 10\,MByte payload. 
In our experiments never more than 400 elephants are existing concurrently, which can easily be handled by HybridTE running on a single machine.
Due to the small number of elephant flows, the flow installation rates on the switches are very low.

Our results show that Hedera does \emph{not} outperform ECMP in terms of the average flow completion time. This result is in line
with what was stated in the original Hedera paper when considering realistic traffic. Thus, in the following discussion we omit Hedera.
In contrast to HybridTE, Hedera is not informed about elephants; it classifies elephants from information gathered from the switches.
We suspect that this classification does not work well with the traffic used in our evaluation. 
Otherwise, if Hedera had full information about all elephants, we would suspect it to perform at least as well as \hte{100}. 
However, even then (due to its high flow installation rate), Hedera would be unusable on real hardware.

\begin{figure}
	\begin{center}
		\includegraphics[width=0.49\textwidth]{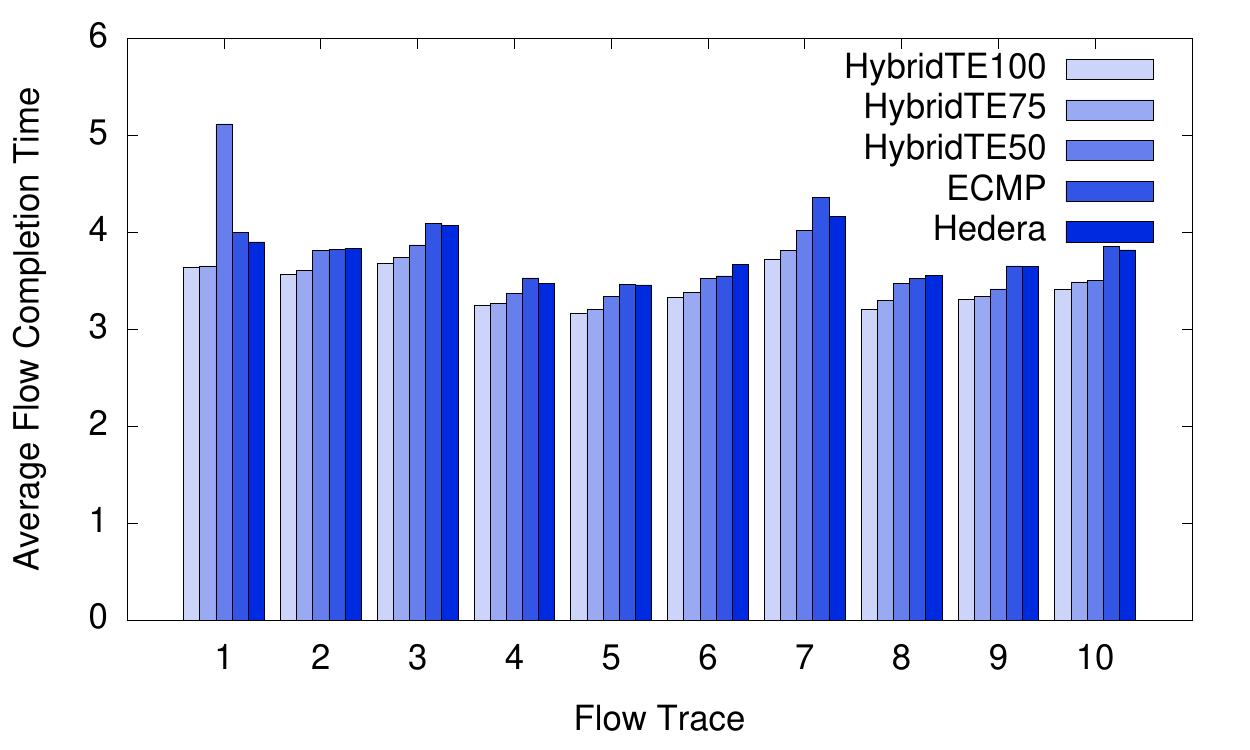}
	\end{center}
	\caption{Average flow-completion time over all ten flow traces on load level 1.}
	\label{fig:flowCompTimeLoad1}
\end{figure}

The average flow completion time in load level 1 can be seen in Figure~\ref{fig:flowCompTimeLoad1}.
For each experiment, the average was computed from the approx. 3 million Layer~4 flows contained in each flow trace.
It can clearly be seen that HybridTE achieves lower flow completion times than ECMP and Hedera. 
The flow completion time resulted from ECMP is on average 7\% higher than for \hte{100}.
\hte{100} and \hte{75} have nearly the same performance: On average, \hte{75} yields 1\% higher flow completion times than \hte{100}.
The difference between \hte{100} and \hte{50} is 6.2\%.
But still, for most flow traces the flow completion times resulted from \hte{50} are significantly lower than from ECMP.

\begin{figure}
	\begin{center}
		\includegraphics[width=0.40\textwidth]{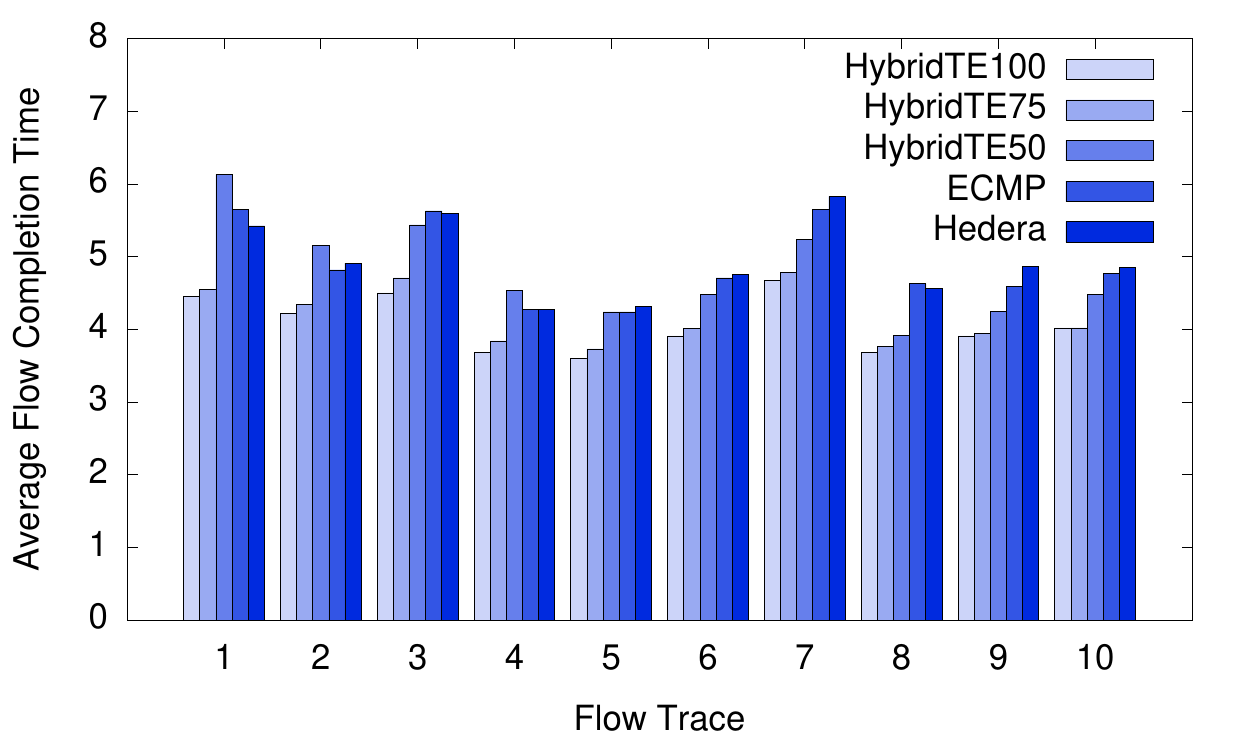}
	\end{center}
	\caption{Average flow-completion time over all ten flow traces on load level 1.5.}
	\label{fig:flowCompTimeLoad15}
\end{figure}

The flow completion times for load level 1.5 are depicted in Figure~\ref{fig:flowCompTimeLoad15}. 
In this load level the emulated network is already congested, which increases flow completion times.
In turn, the results created by ECMP are getting worse: on average ECMP yields 20\% higher flow completion times than \hte{100}. 
We see that with increasing knowledge about elephants, the results are getting better: the gap between \hte{100} and \hte{50} increases from 6.2\% (load level 1) to 14.3\% (load level 1.5).

\begin{figure}
	\begin{center}
		\includegraphics[width=0.40\textwidth]{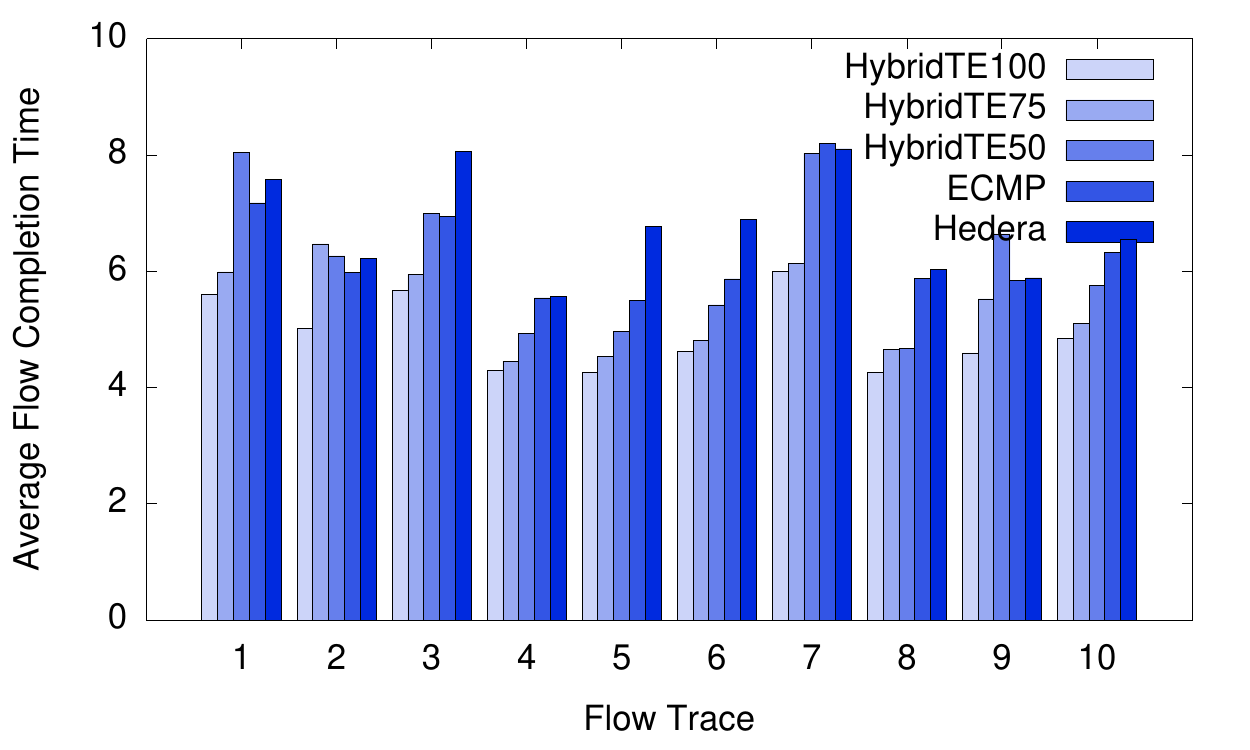}
	\end{center}
	\caption{Average flow-completion time over all ten flow traces on load level 1.75.}
	\label{fig:flowCompTimeLoad175}
\end{figure}
\begin{figure}[!t]
	\begin{center}
		\includegraphics[width=0.40\textwidth]{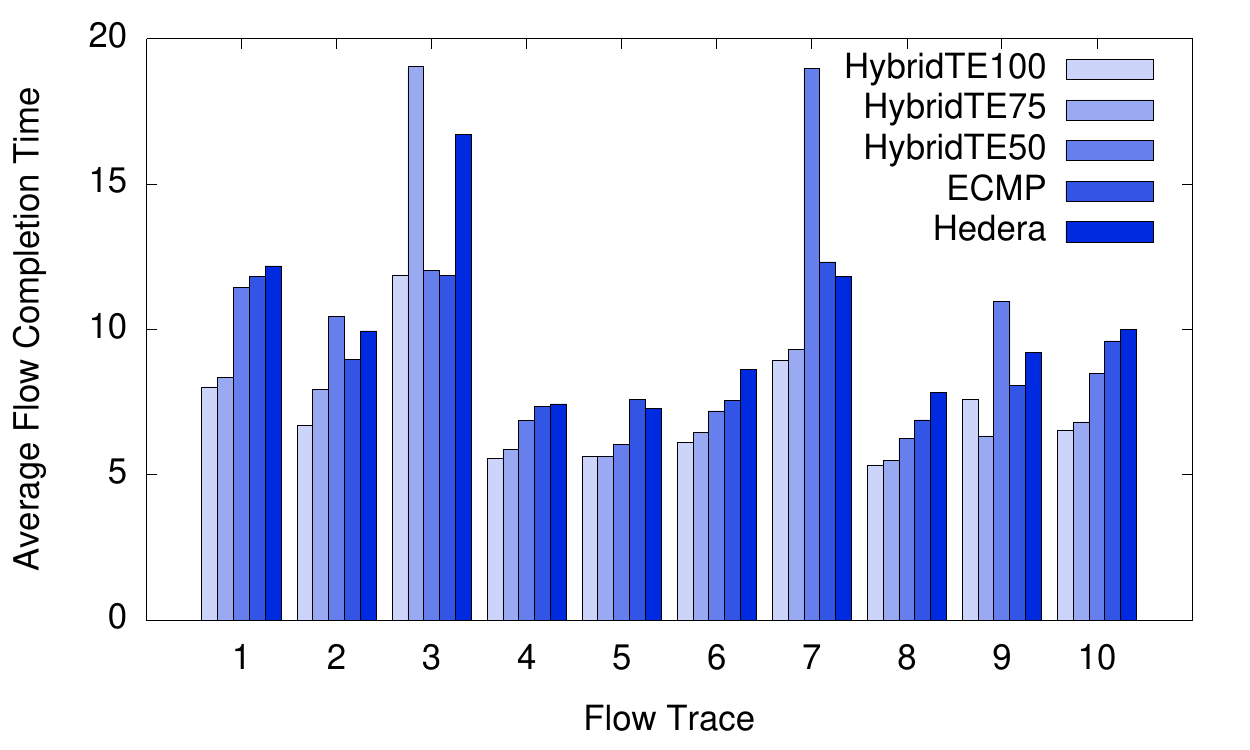}
	\end{center}
	\caption{Average flow-completion time over all ten flow traces on load level 2.}
	\label{fig:flowCompTimeLoad2}
\end{figure}

Figure~\ref{fig:flowCompTimeLoad175} and Figure~\ref{fig:flowCompTimeLoad2} show the results for load levels 1.75 and 2. 
With further increasing congestion in the network, the performance difference between HybridTE and ECMP gets larger. 
At load level 1.75, ECMP yields 28.7\% longer flow completion times, and at
load level 2, the gap even increases to 29.1\%.
At the same time the gap between \hte{50} and ECMP decreases; it seems that in such a heavily loaded network, information about only 50\% of the elephants 
is not enough to provide significantly better results than ECMP.

In summary, our results show that a) with an increasing amount of information, HybridTE yields better results, and b) \mbox{HybridTE} clearly outperforms ECMP and Hedera in terms of flow completion times.
Even with 50\% false negatives, HybridTE still yields better results than ECMP and Hedera.

\subsubsection{False Positives: Mice reported as elephants}
As stated in Section~\ref{sec:isElephantDetectionRealistic}, elephant detection techniques sometimes falsely report mice as elephants.
Whenever a mouse is reported to HybridTE, it will be assigned an individual path through the network. As the mouse only transfers a few bytes it will 
either already have finished when the individual route is installed on the switches or it will use the new route and
complete quickly.
As a result, the newly installed route will time out quickly, too.
Although handling more elephant reports will increase the workload on the controller running HybridTE, 
we do not see any reason for serious performance degradation as long as switch tables do not fill up and flow installation rates stay manageable.
If, however, the rerouting phase of HybridTE starts between the report of a false positive
and its timeout, then this false positive will be part of the input to the rerouting phase and possibly lead to worse rerouting decisions.

\begin{figure}
	\begin{center}
		\includegraphics[width=0.39\textwidth]{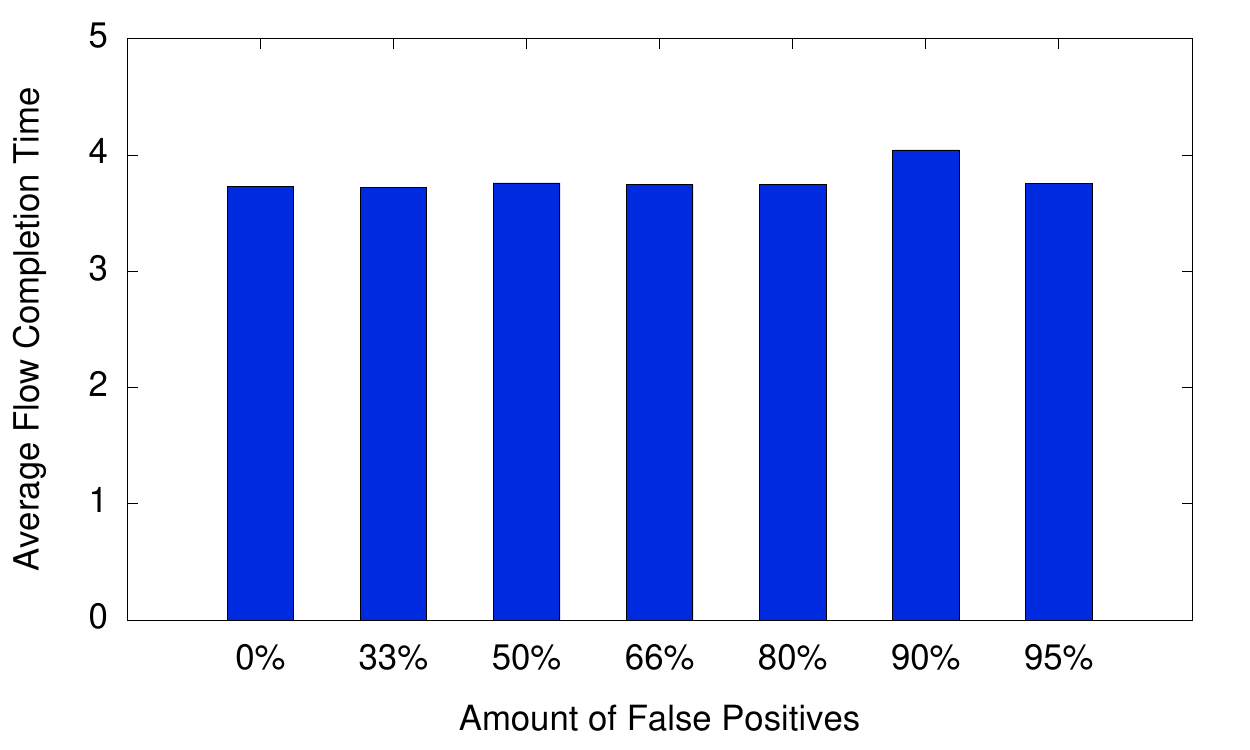}
	\end{center}
	\caption{Average flow completion time created by \hte{100} under different levels of false positive elephant reports on flow trace 7.}
	\label{fig:falsePositives}
\end{figure}

To determine the impact of false positives on the quality of \hte{100}, we conducted the following experiment: 
We fixed the load level to 1 and varied the number of false positives from 0\,\% to 95\,\% where false positives are drawn uniformly at random from all mice.
Figure~\ref{fig:falsePositives} depicts the outcome of this experiment on \emph{flow trace 7}. 
We decided for this particular flow trace because it produced the most distinct results in the previous experiment (Figure~\ref{fig:flowCompTimeLoad1}).
It can clearly be seen that HybridTE is resilient against false positives, which is mainly because
only very few of the false positives live long enough to be rerouted in the first place.

\subsubsection{Delays of Elephant Detection}
\label{sec:heatmap}
In the experiments shown in Section~\ref{sec:results}, HybridTE was informed about elephant flows right when they started. 
In reality, however, the information that a flow is an elephant
will likely be reported to HybridTE during the lifetime of the elephant. This way, in the first phase of its lifetime the elephant it will be handled
by static routing and after some time it will be reported and rerouted, if necessary. We are now going to study the influence of this delay 
on the performance of HybridTE. 
We fixed the load level to 1 and varied two parameters: a) the percentage of false negatives and 
b) the delay between starting an elephant and reporting it to HybridTE.

\begin{figure}
	\begin{center}
		\includegraphics[width=0.38\textwidth]{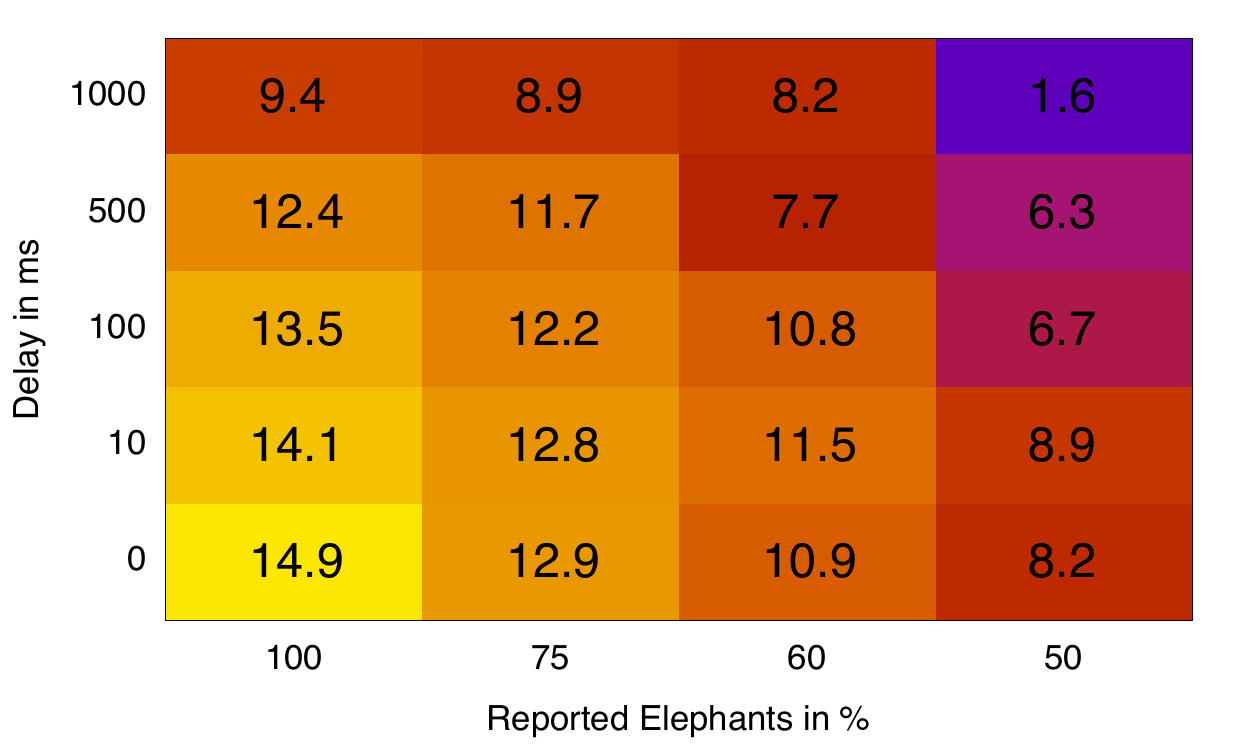}
	\end{center}
	\caption{Reduction (in percent) of the average flow completion time created by HybridTE relative to ECMP under different delays and 
	false negatives.}
	\label{fig:isoPlotDelay}
\end{figure}

The results depicted in Figure~\ref{fig:isoPlotDelay} show the average flow completion time of one run for each parameter tuple on flow trace 7. 
As we did not do any repetitions, it is not enough to make general statements. However, it is enough to get an idea on how the combination between delay 
and the number of reported elephants relates to the performance of \mbox{HybridTE}.
The figure shows the reduction of the average flow compleption time created by HybridTE relative to ECMP.
The delay shown on the y-axis is without time dilation, i.e., a value of 1000\,ms was translated to 150\,s for our emulation.
These experiments confirm the statement of the previous section: HybridTE performs better with an increasing number of reported elephants.
As expected, with increasing delay between starting and reporting an elephant to HybridTE, the performance decreases.

We assume that when using packet sampling techniques, the amount of reported elephants will be between 75\,\% and 100\,\%
while the delay will be between 100\,ms and 1000\,ms. This is a reduction of the average flow completion time between 13.5\,\% and 8.9\% 
compared to ECMP
which is pretty close to \hte{100} when reporting elephants right away.

\section{Conclusion}
\label{sec:conclusion}
We showed that when using information from elephant detection techniques, we can construct a simple routing algorithm that is very efficient in 
resource usage while keeping the same (or even better) performance as state-of-the-art TE techniques leveraging uncertain information about
elephant flows.
Moreover, we have shown that our approach is extremely robust against false positives and can withstand even adverse false 
negatives of 50\,\% and elephant reporting delays of up to one second.
Using this algorithm one can construct SDN networks using low-cost data-center hardware with similar performance characteristics as networks from
expensive end-of-line equipment.
	 
\vspace{-0.3cm}
\section*{Acknowledgments}
\vspace{-0.13cm}
\begin{small}
	This work was partially supported by the German Research Foundation (DFG) within the Collaborative Research Centre ``On-The-Fly Computing'' (SFB 901).
\end{small}

\newpage

\bibliography{hybridTE}{}
\bibliographystyle{unsrt}

\balancecolumns
\end{document}